# Towards Novel Malicious Packet Recognition: A Few-Shot Learning Approach

Kyle Stein[1], Andrew A. Mahyari[1,2], Guillermo Francia, III[3], Eman El-Sheikh[3]
[1] Department of Intelligent Systems and Robotics, University of West Florida, Pensacola, FL, USA
[2] Florida Institute For Human and Machine Cognition (IHMC), Pensacola, FL, USA
[3] Center for Cybersecurity, University of West Florida, Pensacola, FL, USA
ks209@students.uwf.edu, amahyari@uwf.edu, gfranciaiii@uwf.edu, eelsheikh@uwf.edu

*Abstract*—As the complexity and connectivity of networks increase, the need for novel malware detection approaches becomes imperative. Traditional security defenses are becoming less effective against the advanced tactics of today's cyberattacks. Deep Packet Inspection (DPI) has emerged as a key technology in strengthening network security, offering detailed analysis of network traffic that goes beyond simple metadata analysis. DPI examines not only the packet headers but also the payload content within, offering a thorough insight into the data traversing the network. This study proposes a novel approach that leverages a large language model (LLM) and few-shot learning to accurately recognizes novel, unseen malware types with few labels samples. Our proposed approach uses a pretrained LLM on known malware types to extract the embeddings from packets. The embeddings are then used alongside few labeled samples of an unseen malware type. This technique is designed to acclimate the model to different malware representations, further enabling it to generate robust embeddings for each trained and unseen classes. Following the extraction of embeddings from the LLM, few-shot learning is utilized to enhance performance with minimal labeled data. Our evaluation, which utilized two renowned datasets, focused on identifying malware types within network traffic and Internet of Things (IoT) environments. Our approach shows promising results with an average accuracy of 86.35% and F1-Score of 86.40% on different malware types across the two datasets.

## I. INTRODUCTION

As the digital world expands with the sophistication of networks and devices, the urgency for robust malware detection intensifies. With the constant evoluton of malware being introduced in cyberattacks, traditional security mechanisms, such as firewalls and antivirus software, are becoming less effective. These traditional defenses mainly depend on recognizing predefined packet headers and known malware signatures, which leaves a gap for more complex cyberattacks that do not match existing and previously learned patterns. With an estimated 560,000 new malware instances identified daily, and over 1 billion malware programs currently known [1], the sheer volume and diversity of these threats necessitate a shift from traditional signature-based detection methods to systems that are more adaptive and intelligent. Deep Packet Inspection (DPI) steps in as a crucial technology for analyzing network traffic in detail, surpassing standard network monitoring tools by inspecting not just the packet's header information, like source and destination IP addresses, but also scrutinizing the data payload. DPI's integration with machine learning algorithms can further enhance its effectiveness, allowing for the dynamic identification of novel threats by analyzing patterns in network payloads, thereby bridging the gap left by traditional security measures and fortifying defenses against introduced cyber threats. This analysis provides a deeper insight into the traffic's nature, allowing for the detection of malicious activities that might otherwise remain hidden.

In the context of advancing malware detection within networks, several related works underscore the shift towards more intelligent and adaptive security solutions. In [2], researchers highlighted the application of deep learning techniques for enhancing DPI capabilities, specifically using convolutional neural networks (CNNs) and fully connected neural networks to analyze and classify network traffic. Xu, et al. [3] investigated the efficacy of ensemble learning models for malware detection in networks, emphasizing the importance of developing models capable of swiftly adapting to novel malware signatures. Farrukh, Wali, Khan and Bastian [4] employed a method where packet-level byte data was converted into image-based representations. These images were then used to train two types of datasets: a base learner dataset containing examples of benign data, and a meta learner dataset comprising instances of known cyber attacks. This training enabled meta-classifier models to identify unknown attacks. Stein, Mahyari, and El-Sheikh [5] showed that payloads which are present in the Controller Area Network (CAN) of vehicles can be classified as either benign or malicious based on an end-to-end Recurrent Neural Network (RNN). However, it is important to note the fundamental differences between the structure of CAN and IP network payloads. The CAN bus supports a maximum message payload of eight bytes per frame, while the typical maximum transmission unit (MTU) of an IP network packet is 1500 bytes.

In this study, we propose a novel framework for detecting and classifying malware. In this approach, our method relies on incorporating a LLM-based architecture [6], capitalizing on the self-attention mechanism to decipher the raw bytes within network packets effectively. Our proposed approach leverages the capabilities of LLMs to learn the sequential patterns of packets and extract embeddings from packets that best represent them. The main **contribution of our approach** lies in learning and classifying novel types of malware from a limited set of labeled samples (i.e. few shots). This enables the transfer of knowledge learned from the known malware types to the newly introduced malware with only few labeled examples [7], enhancing our model's ability to quickly adapt and recognize novel malware threats with limited exposure. This few-shot adaptability, combined with the LLM's focus on producing class-specific embeddings, elevates our classification methodology beyond traditional malware detection systems by offering a robust solution to the advancement of introduced malware threats.

## II. DATA AND PRE-PROCESSING

In this section, we describe the pre-processing steps to prepare the datasets for analysis. We discard packets that do not contain any payloads, corresponding to handshakes, acknowledgment, and any other network protocols and only focus on the packets that contain payloads.

### A. UNSW-NB15

The UNSW-NB15 dataset [8] was developed for enhancing public intrusion detection datasets, containing 100 GB of raw network data in PCAP format with a mix of real and simulated attacks. It includes comprehensive ground truth labels, the network five-tuple, and attack timelines for in-depth network analysis. We utilize the extracted UNSW-NB15 dataset made available in [11]. The dataset preparation involves extracting five-tuple information from TCP/UDP packets, converting packet bytes to hexadecimal, and removing duplicates to form a detailed matrix of network interactions. This matrix is then cross-referenced with ground truth labels, based on the IP addresses, ports, and attack timings, to classify traffic packets as benign or malicious accurately. For the model, packet bytes, converted to decimal integers, are strictly the main input. We balance the dataset by selecting equal numbers of malicious packets between three chosen attacks and are subsequently labeled. It is important to randomly select an equal amount of malicious entries to ensure the model is not biased toward the majority class, which could lead to a significant number of false positives or negatives. The dataset is further refined by removing duplicate packet bytes to also prevent bias. For multi-class analysis, three attack types - Fuzzers, Exploits, and Generic - are focused on due to their payload sophistication, differentiating from simpler, volume-based attacks like DDoS.

### B. CIC-IoT23

The CIC-IoT23 [9] is a comprehensive dataset specifically tailored to the study of Internet of Things (IoT) security vulnerabilities. It encompasses seven distinct classes of cyberattacks, including Distributed Denial of Service (DDoS), Brute Force, Spoofing, Denial of Service (DoS), Recon, Web-based, and Mirai attacks. Furthermore, these seven classes are broken down into over thirty specific attacks against IoT devices. This dataset is notable for its detailed representation of network traffic gathered from a robust IoT setup comprising 105 distinct devices. It provides a rich compilation of PCAP files that record both benign and malicious network activities. The pre-processing approach for the CIC-IOT23 dataset is similar to that of the UNSW-NB15, starting with the extraction of transport layer bytes from all TCP or UDP packets within the selected four PCAP files. The extracted packets bytes are converted to hexadecimal, with duplicates removed to ensure data quality.

After converting the packet bytes first to hexadecimal and then to decimal integers, each malicious class is given a respective label. To maintain a balanced dataset and mitigate model bias, an equal number of malicious entries from each class are randomly chosen. Similarly to the UNSW-NB15 dataset, our study specifically focuses on three types of attacks for the CIC-IOT23: Backdoor Malware, Vulnerability Attack, and Brute Force Attack. These attacks rely on the specific bytes within the payload to carry out their malicious goals, similar to the attacks chosen from the UNSW-NB15 dataset.

## III. METHODOLOGY

In this section, we describe the system architecture, training, and evaluation of our model for deep packet inspection. Leveraging LLM's capability to capture contextual patterns, we show that it can be effectively applied to discern distinctions in packet sequences. This process is powered by the self-attention mechanism [12], enabling each byte in the input sequence to reference other bytes within that sequence. The computation of attention weights for each byte pair highlights the importance of each byte, informing the model's understanding and processing of the sequence. The architecture of the LMM is comprised of several key components:

**Embedding Layer:** This layer acts as the gateway for input data into the LLM framework, converting the input sequence, $\mathbf{X} = (x_1, x_2, ..., x_N)$, into a matrix representation, $\mathbf{E} \in \mathbb{R}^{N \times d_{model}}$, where $N$ is the number of decimal integer values in the sequence and $d_{model}$ is the dimension of the embedding vectors. Through the embedding matrix $\mathbf{W}_E$, the input $\mathbf{X}$ is mapped to the feature space, resulting in $\mathbf{Z}$, a matrix representation of the input features.

$$\mathbf{Z} = \mathbf{W}_E \mathbf{X} \tag{1}$$

**Positional Encodings:** To provide the LLM with an understanding of sequence order, positional encodings are added to the embeddings. This encoding assigns a unique identifier to each sequence position, allowing the model to recognize and utilize the order of bytes in a packet. The encoding for each position $pos$ and dimension $i$ is computed as follows:

$$PE_{pos,2i} = \sin\left(\frac{pos}{10000^{2i/d_{model}}}\right) \tag{2}$$

$$PE_{pos,2i+1} = \cos\left(\frac{pos}{10000^{2i/d_{model}}}\right) \tag{3}$$

**Multi-head Self-Attention Mechanism:** This mechanism enables the model to evaluate and integrate information across the entire sequence. It transforms the input sequence $\mathbf{Z}$ into three distinct sets of vectors—queries ($\mathbf{Q}$), keys ($\mathbf{K}$), and values ($\mathbf{V}$)—using learned weight matrices. These vectors facilitate the computation of attention weights:

$$\mathbf{Q} = \mathbf{W}_Q \mathbf{Z} \tag{4}$$
$$\mathbf{K} = \mathbf{W}_K \mathbf{Z} \tag{5}$$
$$\mathbf{V} = \mathbf{W}_V \mathbf{Z} \tag{6}$$

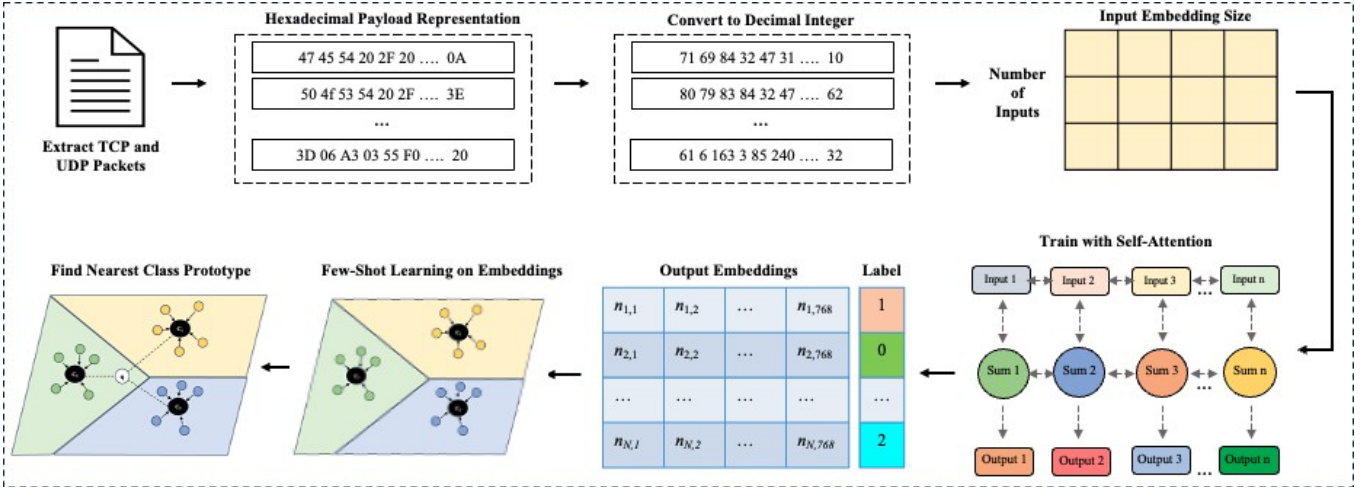

Fig. 1: **The proposed packet detection algorithm processes packet byte data by initially converting it to hexadecimal and then to decimal integer values between [0, 255]. The LLM is designed to train on various combinations of classes, using different permutations of malwares. This approach allows the model to generate a unique embedding matrix for each class combination. Following the transformation of packet data through the self-attention mechanism, the output embeddings are then leveraged in a few-shot learning framework. Within this framework, the embeddings from the query set are classified against those in the support set by determining their closest match through Euclidean distance. The assignment of the appropriate class label is based on the proximity to the nearest class prototype.**

Attention scores are calculated via a scaled dot-product of queries and keys, normalized using the softmax function, to guide the aggregation of values:

$$\text{Attention}(\mathbf{Q}, \mathbf{K}, \mathbf{V}) = \text{softmax}\left(\frac{\mathbf{Q}\mathbf{K}^T}{\sqrt{d_k}}\right)\mathbf{V} \quad (7)$$

This mechanism allows the model to capture both the global understanding and complex information necessary for understanding the packet input patterns.

**Output Layer and Embedding Extraction:** The output layer's primary role is to derive embeddings, **E**, which encapsulate the learned patterns for each sequence. These embeddings are averaged across the sequence to distill the contextual information, offering a condensed yet rich representation of the original data, suitable for detailed analysis or classification in downstream processes. This application of our LLM for classifying network traffic, including the essential steps from initial embedding to few-shot learning with prototypical networks, is detailed in Algorithm 1.

### A. Model Configuration, Training, and Fine-Tuning

The selected hyperparameters of the model are critical for its performance and were chosen based on the specific requirements of our application. The model incorporates a diverse set of parameters, including 256 unique bytes, which signifies the total number of distinct elements identified within the input data. This set accounts for each unique hexadecimal value alongside an additional padding integer, ensuring comprehensive coverage of the input space. The hidden size is set at 768, which determines the dimensionality of the hidden states and is pivotal for the model's capacity to process and represent information. The architecture is further defined by 12 hidden layers, establishing the depth of the network and enhancing its ability to learn complex patterns within the data. Accompanying these layers, 12 attention heads are employed to facilitate the model's focus on various segments of the input simultaneously, a crucial aspect for understanding and learning from the data efficiently. The intermediate size is configured to 3072, which specifies the dimension of the feed-forward layers within each transformer block, playing a vital role in the network's ability to transform input data at each stage effectively. The maximum position embeddings are set to 1500, corresponding to the maximum length of the input sequence that the model can handle. This configuration allows the model to process input sequences without imposing a strict limit on the packet length, thereby accommodating a wide range of data inputs.

The model is trained over 15 epochs, utilizing cross-entropy as the foundational loss function. Optimization is conducted using the Adam optimizer, configured with a learning rate of $2e-5$, and incorporates weight decay regularization to curb excessive weight growth, aiding in the prevention of overfitting [13]. Additionally, a learning rate scheduler is employed to modulate the learning rate throughout the training duration. This scheduler progressively increases the learning rate from zero up to a predetermined maximum during the warm-up phase, followed by a linear reduction of the learning rate across the subsequent training epochs [14]. We train our model on various combinations of known classes, using permutations of class pairs at any given time. This strategy is pivotal for generating a robust set of embeddings from the LLM, which few-shot learning then leverages for classification. This approach not only enhances our model's adaptability to new malware types but also deepens its understanding of subtle distinctions between malware classes. The model was trained using NVIDIA GeForce RTX 2080 GPUs.

**Algorithm 1** Proposed Framework for Network Traffic Classification

**Input:** Set of network packet bytes $P = \{p_1, p_2, \ldots, p_n\}$
**Output:** Classifications for network traffic: Malware Type Identification
**Step 1: LLM Model Training**
Embed packets $X$ using an embedding matrix $\mathbf{W}_E$ to get $\mathbf{Z} = \mathbf{W}_E \mathbf{X}$
Add positional encodings $\mathbf{PE}$ to $\mathbf{Z}$
Train the LLM with self-attention, optimizing with cross-entropy loss and the Adam optimizer
**Step 2: Embedding Extraction**
Extract embeddings $\mathbf{E}$ from the LLM's final layer for downstream tasks
**Step 3: Few-Shot Learning with Prototypical Networks**
Perform episodic training selecting $K$ classes, support $S_k$ and query $Q_k$ sets, and number of classes per episode $N_c$
Compute class prototypes $c_k = \frac{1}{N_C} \sum_{(x_i, y_i) \in S_k} f_\phi(x_i)$
Classify queries by nearest prototype and update model with loss $J$

### B. Few-Shot Learning Architecture

In this section, we describe how few-shot learning is used to transfer a pretrained LLM model on known malware types to new malware types. The selection of this approach is motivated by its efficacy in learning accurately from a limited number of labeled examples, a critical feature for cybersecurity domains facing rapid malware evolution. Models capable of quick adaptation with few examples are essential.

To develop the few-shot learning algorithm, we use a pretrained LLM to extract packet embeddings. The LLM is pretrained from scratch on known malware types. We first define the terminology used in our few-shot learning algorithm. Support set consists of a limited number of labeled examples from each class, while the query set contains unlabeled examples that the model attempts to classify based on the knowledge learned from the support set. The model uses a limited number of support examples (shots) from each class (ways) to form class-specific prototypes. These prototypes are the mean representation of each class based on the embeddings. The architecture incorporates a fully connected linear layer to transform the input embeddings of packets (extracted using the pretrained LLM) into a suitable representation space for classification. The model undergoes episodic training, a technique that structures training into small, problem-specific episodes, enhancing its ability to generalize during actual deployment.

Algorithm 1 outlines the process for differentiating between malicious network traffic using episodic training. In each episode, $K$ classes are randomly selected from the dataset. For each class $k$, a support set $S_k$ with $N_S$ examples and a query set $Q_k$ with $N_Q$ examples are prepared. Class prototypes $\mathbf{c}_k$ are computed as the mean of the support set embeddings' transformed feature vectors from the network's last layer. Next, the algorithm evaluates each query example $(x, y)$ in $Q_k$, calculating its Euclidean distance to each class prototype $\mathbf{c}_k$. The example $x$ is then assigned the label of the closest prototype. Model updates are based on the loss $J$, calculated from the softmax probabilities of the distances between query examples and prototypes, shown in the equation below.

$$J \leftarrow J + \frac{1}{N_C N_Q} \left[ d(f_\phi(x), c_k) + \log \sum_{k'} \exp(-d(f_\phi(x), c_{k'})) \right] \quad (8)$$

This iterative process across numerous episodes aims to improve the model's generalization to new class examples.

## IV. RESULTS AND DISCUSSION

To evaluate the performance of the proposed method, we consider accuracy and F1-Score on the query dataset. It is important to note that the query dataset used in our evaluation process differs from the support dataset. The query dataset consists of randomly selected, unseen samples and ensures that the model is using a distinct dataset to evaluate the performance. This evaluation includes tests for 3-way classification scenarios, utilizing setups with 5 and 10 shots along with 15 query samples. The query samples are composed of instances from each class, rather than solely focusing on the untrained class. This choice ensures our model maintains a full understanding towards all classes, not just the novel class. By incorporating samples from both the trained and untrained classes in the query set, we aim to present a balanced learning environment for the model to adapt to all classes.

For a comprehensive evaluation, we conduct this training process over 10 epochs, where each epoch consists of 1,000 and 5,000 episodes for the UNSW-NB15 and CIC-IoT23 datasets, respectively. This variation in the number of episodes corresponds with the size differences between the balanced datasets: UNSW-NB15 contains more than 8,000 samples, while CIC-IoT23 has over 40,000 samples. In our empirical study, we determined that the respective number of episodes per epoch balanced maximum learning and computational efficiency. This configuration also allowed the model to adequately learn from the data without overfitting, considering the size and complexity differences between the datasets. To assess the robustness of the model, we repeat the entire training and evaluation process across 10 iterations. Each iteration involves re-initializing the model's weights and undergoing the training procedure from scratch, spanning all epochs and episodes. This approach allows us to gather a more consistent average on the model's performance, providing a more reliable measure of the model's ability to learn from limited examples and adapt to new, unseen data.

The results for few-shot learning to classify new and old types of malwares are shown in Table I. The 'Trained Classes' column shows which classes the LLM was trained on, and subsequently which class was not trained for that experiment. The class which is not trained is the introduced class (or novel class) for that respective experiment. Our few-shot learning approach transfers the knowledge learned from the known, or trained, malware types to a newly introduced novel type of malware with a limited amount of labeled samples. The best performance on the UNSW-NB15 dataset occurred when the model was trained on Fuzzers and Generic malwares, with Exploits being introduced, correctly classifying 92% of query samples from all three classes. For the CIC-IoT23 dataset, training on Vulnerability and Brute Force attacks, with Backdoor attacks being introduced, led to an 88% accuracy on the query samples. We extend the assessment to compare the overall performance of our proposed method in relation to other well-established methodologies [2] [3] shown in Tables II and III. Unlike these models, which underwent end-to-end training on all three classes, our method achieves similar or superior performance despite utilizing significantly less labeled data and training on only two out of the three classes. For the results presented in the tables, we calculated averages for both 5 and 10 shot scenarios.

TABLE I: Results of Few-Shot Classification of Malware Types on the Query Set

| Dataset | Trained Classes | Proposed Method | | | |
| --- | --- | --- | --- | --- | --- |
| | | 5-shot | | 10-shot | |
| | | Acc. | F1 | Acc. | F1 |
| UNSW-NB15 | Exploits, Fuzzers | 88.11 | 88.11 | 88.22 | 88.22 |
| | Fuzzers, Generic | 89.11 | 89.06 | **92.03** | **92.03** |
| | Exploits, Generic | 84.89 | 84.93 | 86.22 | 86.23 |
| CIC-IoT23 | Backdoor, Vulnerability | 83.11 | 82.91 | 83.33 | 83.23 |
| | Vulnerability, Brute Force | 82.22 | 82.40 | **88.00** | **88.79** |
| | Backdoor, Brute Force | 86.89 | 86.96 | 84.00 | 83.87 |

TABLE II: Comparison of Methods on UNSW-NB15 Dataset

| Method | Accuracy | F1-Score |
| --- | --- | --- |
| 1D-CNN [2] | 86.65 | 86.80 |
| 2D-CNN [3] | 87.90 | 88.06 |
| LSTM [3] | 75.68 | 76.14 |
| Proposed Method (Average) | **88.10** | **88.10** |

TABLE III: Comparison of Methods on CICIOT-23 Dataset

| Method | Accuracy | F1-Score |
| --- | --- | --- |
| 1D-CNN [2] | 84.37 | 84.29 |
| 2D-CNN [3] | 82.53 | 82.41 |
| LSTM [3] | 82.52 | 82.45 |
| Proposed Method (Average) | **84.59** | **84.69** |

The main idea in comparing a LLM with other deep learning approaches is to examine different methods for processing sequences. Although each of these models deal with sequences at their core, they do so through unique methods. Convolutional Neural Networks (CNNs) are adept at identifying localized patterns and layered structures within the data, making them particularly useful for detecting patterns within segments of network traffic. Long Short-Term Memory (LSTMs) leverage recurrent connections to retain information across lengthy sequences. However, our model excels at understanding and identifying context across broader spans without the iterative approach of LSTMs, thus avoiding potential data loss. The self-attention mechanism enables the model to assess the importance of various parts of the sequence, facilitating a global-understanding of the information.

Our research stands out in several key ways. We began by removing duplicated packets, enhancing the model's exposure to distinct byte data and thereby improving data quality for more effective generalization to new data. Retaining these duplicates can reduce the uniqueness of signatures, making the process for classification algorithms more trivial in identifying which packets belong to a certain class. Our analysis was not limited to the initial 784 bytes as described in [2], but extended to a maximum of 1500 bytes. Given that the typical Maximum Transmission Unit (MTU) for Ethernet packet-based networks is 1500 bytes, this approach ensures our model considers the full length of available data. This is important as malicious content might be embedded further within a packet's payload. This method is also effective in a multi-class classification scenario, whereas some previous malware detection efforts are primarily concentrated on binary classifications [16], [17].

## V. ENCRYPTED TRAFFIC

In this study, our analysis focuses on the raw packet bytes contained within TCP and UDP network packets from various datasets. However, it is crucial to note the limitations when dealing with encrypted traffic. Encryption—the process of converting plaintext into ciphertext to secure data transmission and make it unreadable to unauthorized entities without the appropriate decryption keys—plays a pivotal role in data security. Specifically, when plaintext is encrypted using a key, it transforms into ciphertext that does not reveal any information about the original content without the correct decryption key [18].

Let us denote $m$ as the variable representing the raw packet message, and $E_k(\cdot)$ as the encryption function that encrypts $m$ with the key $k$. In this context, observing the ciphertext $c$ yields no information about the original message $m$, as:

$$p(m = m_1 | E_k(m) = c) = p(m = m_1) \qquad (9)$$

This equation underscores that decrypting the plaintext $m_1$ with the encryption algorithm $E_k(\cdot)$ and key $k$ results in a unique ciphertext, effectively concealing any details of the plaintext.

Despite these encryption properties, some studies have successfully classified network traffic from various applications based off encrypted packets [19], [20]. This classification success is largely due to the unique encryption signatures that different applications' random generators leave, which, while not disclosing plaintext content, offer a form of application-specific pattern [21]. The challenge, however, lies in malware detection, where access to plaintext information is crucial. If malware signatures can be detected in encrypted traffic, it would suggest that the encryption method fails to adequately mask the plaintext, thereby violating the principle stated in Equation (9).

This research involved applying two encryption methods on packet data: the Advanced Encryption Standard (AES)

and Fernet symmetric encryption. AES uses a 256-bit key and a 16-byte initialization vector (IV) to encrypt the raw data, ensuring a unique encryption outcome for each key and IV pair [22]. Fernet employs AES in Cipher Block Chaining (CBC) mode with a 128-bit key for encryption [23]. Following the encryption, we processed them according to the pre-processing methodology described earlier, converting each encrypted packet byte into a sequence of decimal integers and then padding these sequences to ensure uniform length for each input. These inputs were then trained and tested using the method's architecture from [24].

The AES encryption algorithm yielded an average test accuracy of 57.16% and an F1-Score of 44.52%, indicating that the model was unable to effectively distinguish between malicious and benign packets. This suggests that AES encryption successfully diminishes the learnable patterns within the raw bytes, which our model relies on for packet classification. In contrast, the Fernet encryption algorithm achieved a test accuracy of 91.41% and an F1-Score of 92.09%, indicating that not all encryption algorithms are equally effective at concealing plaintext information, while others can.

## VI. Conclusion and Future Work

This paper explored the application of a LLM and few-shot learning for malware classification. The model was evaluated on the UNSW-NB15 and CIC-IDS2017 datasets, focusing on the packet bytes of UDP and TCP packets serving as inputs. Our method showed robust results on the classification of different malware types with limited amounts of training data when compared to state-of-the-art methods. While the structure of packet bytes are significantly different from natural language, this study shows that the LLM developed for natural language can be leveraged to capture and learn the intricate sequential patterns of the packet bytes. Future work in this domain will include applying self-supervised techniques to help further learn the intricacies of packet bytes, distinguishing between packet header and payload bytes, and exploring the potential for cross-dataset generalization to enhance model robustness across various network environments.

## VII. Acknowledgement

This work is partially supported by the UWF Argo Cyber Emerging Scholars (ACES) program funded by the National Science Foundation (NSF) CyberCorps® Scholarship for Service (SFS) award under grant number 1946442. Any opinions, findings, and conclusions or recommendations expressed in this document are those of the authors and do not necessarily reflect the views of the NSF.